\documentclass[12pt]{article}
\usepackage{graphicx}

\def\la{\;\raise0.3ex\hbox{$<$\kern-0.75em\raise-1.1ex\hbox{$\sim$}}\;}
\def\ga{\;\raise0.3ex\hbox{$>$\kern-0.75em\raise-1.1ex\hbox{$\sim$}}\;}

\def\a{\alpha}
\def\adef{\,\alpha=e^2/\hbar c\,}
\def\mudef{\,\mu=m_p/m_e\,}

\begin{document}
\thispagestyle{empty}
\thispagestyle{empty}

\begin{center}
{\Large\bf Do the fundamental constants vary \\
in the course of the cosmological evolution ? \\}

\bigskip\bigskip\bigskip \bigskip

{\large\it  A.V.~Ivanchik$^1$, E.~Rodriguez$^2$,
P.~Petitjean$^{2,3}$, and D.A.~Varshalovich$^1$}
\par
\bigskip \bigskip \bigskip \bigskip

$^1$ Ioffe Physical Technical Institute RAS, St.-Petersburg, Russia \\
$^2$ Institut d'Astrophysique de Paris -- CNRS, France \\
$^3$ LERMA - Observatoire de Paris, France

\end{center}
\bigskip\bigskip\bigskip \bigskip \bigskip \bigskip \bigskip \bigskip
{\bf Abstract} --
We estimate the cosmological variation of the proton-to-electron mass ratio
$\mudef$ by measuring the wavelengths of molecular hydrogen transitions in
the early universe. The analysis is performed using high spectral resolution
observations ($FWHM\approx 7\;$km/s) of two damped Lyman-$\alpha$ systems at
$z_{abs}=2.3377$ and $3.0249$ observed along the lines of sight to the
quasars Q~1232+082 and Q~0347$-$382 respectively.
\par\noindent
The most conservative result of the analysis is a possible variation of
$\mu$ over the last $\sim$10~Gyrs, with an amplitude
\[\Delta \mu / \mu = (5.7\pm3.8)\times 10^{-5}.\]
The result is significant at the 1.5$\sigma$ level only and should be
confirmed by further observations.
This is the most stringent estimate of a possible cosmological variation of
$\mu$ obtained up to now.
\par\bigskip
\bigskip\bigskip \bigskip \bigskip \bigskip \bigskip \bigskip

\vspace{2cm}$^*$E-mail: iav@astro.ioffe.rssi.ru

\newpage
\section*{Introduction}
\hspace{1cm}
Contemporary theories of fundamental interactions (SUSY GUT,
Superstrings/M-theory and others) predict that fundamental physical constants
change in the course of the Universe evolution. First of all, coupling
constants vary with increasing energy transfer in particle interactions
(corresponding to the so-called ``running constants''). This effect has been
proved by high-energy experiments in accelerators. For example,
the fine-structure constant $\adef$ equals $1/137.036$ at low energies
$(E\rightarrow 0)$, but increases up-to $1/128.896$ for energy
$E\sim 90$~GeV (Vysotsky et al. 1996). Such ``running'' of the constants has
to be taken into account in any consideration of the very early Universe.
Another prediction of the current theories is that the low-energy limits of
the constants can vary during the cosmological evolution and take different
values at different points of the space-time. There are several reasons for
such variations.
In multidimensional theories (Kaluza-Klein, ``p-brane'' models and others)
variations of fundamental physical constants are a direct result of the
cosmological evolution of extra-dimensional sub-spaces. In  some theories
(e.g. Superstrings) variations of the constants are a consequence of the
evolution  of the vacuum state
(a vacuum condensate of some scalar field or ``Quintessence'').

Clearly, experimental detection of such variations of the constants would be
a great step forward in our description of Nature.
Recently,  Webb et al. (2001) announced the detection of a possible variation of
the fine-structure constant, $\Delta \a / \a = (-0.72\pm0.18)\times10^{-5}$,
over a redshift range $0.5<z<3.5$.
The method used by these authors is based on the measurement of the
variation of a large number of transitions from different species.
This decreases significantly the statistical errors.
However, the estimate of the systematics becomes more complicated than
in the method used earlier where only one species was
considered (e.g. Ivanchik et al., 1999).
In any case, this exciting result should be checked using some other method.

\section*{Proton-to-Electron Mass Ratio}
\hspace{1cm} Here we test a possible cosmological variation of
$\Delta \mu / \mu$, where $\mu$ is the proton-to-electron mass
ratio at the epoch $z=2.3-3.0$. It should be noted that a
variation of $\a$, in principle, implies a variation of $\mu$,
because any kind of interaction inherent to the particle gives a
contribution to its observed mass. This means that any variation
of the interaction parameters has to produce some variation of the
particle mass, and consequently $\mu$. Unfortunately, the physical
mechanism generating the masses of the proton and the electron is
unknown up to date. Therefore, the exact functional dependence of
$\mu(\a)$ is unknown too. Nevertheless, there are some models
which permit to estimate the electromagnetic contribution to the
mass of proton and electron (e.g. Gasser \& Leutwyler, 1982) and
dependence of $m_p$, $m_e$, and $\a$ on a scalar field which may
changed during the evolution (e.g. Damour \& Polyakov, 1994).
There are model relations between cosmological variation of $\a$
and $m_p$ (e.g. Calmet \& Fritzsch, 2001). In addition, a curious
numerical relation may be mentioned: the dimensionless constant
$\mudef$ approximately equals the ratio of the strong interaction
constant $g^2/(\hbar c) \approx 14$ to the electromagnetic
interaction constant $\adef \approx 1/137$, where $g$ is the
effective coupling constant calculated from the amplitude of the
$\pi$-meson--nucleon scattering at low energy.

At present, the proton-to-electron mass ratio is measured within
the relative accuracy of $2\times 10^{-9}$ and equals $\mu =
1836.1526670(39)$ (Mohr \& Tailor, 2000). Laboratory metrological
measurements rule out considerable variation of $\mu$ on a short
time scale, but do not exclude changes over the cosmological time,
$\sim 10^{10}$ years. Moreover, one can not reject a priori the
possibility that $\mu$ (as well as other constants) takes
different values in widely separated regions of the Universe. This
should be directly tested by means of astrophysical observations
of distant extragalactic objects. Measurements of the wavelengths
of absorption lines in high-redshift quasar spectra is a powerful
tool to check directly the possible variation of the constants
over the cosmological time (e.g. $\mu$ and $\a$) between present
days and the epoch at which the absorption-spectrum has been
produced, i.e.$\sim 10-13$~Gyr ago.

Up to now the most stringent estimate of possible cosmological
variation of $\mu$ was obtained by Potekhin et al. (1998), viz
$\Delta \mu/\mu < (-10\pm8)\times 10^{-5}$.

\section*{Sensitivity Coefficients}
\hspace{1cm}
The method used here has been suggested by Varshalovich \& Levshakov (1993)
and is based on the fact that wavelengths of electron-vibro-rotational lines
depend on the reduced mass of the molecule. It is essential that this
dependence differs for different transitions.
This makes it possible to distinguish the cosmological redshift of a line
from the shift caused by a possible variation of $\mu$.
The change in wavelength $\lambda_i$ due to variation of $\mu$ may be
described (in the case $\Delta \mu / \mu << 1$) by
{\it the sensitivity coefficient} $K_i$ defined as
\begin{equation}
K_i=\frac{\mu}{\lambda_i}\frac{\mbox{d} \lambda_i}{\mbox{d} \mu}
\end{equation}
Such coefficients were calculated for the Lyman and  Werner bands of
molecular hydrogen by Varshalovich \& Levshakov (1993),
and Varshalovich \& Potekhin (1995). Thus, the measured wavelength
$\lambda_i$ of a line formed in the absorption system at redshift $z_{abs}$
can be written as
\begin{equation}
\lambda_i=\lambda_i^0\cdot(1+z_{abs})\cdot(1+K_i \cdot \Delta\mu/\mu) \, ,
\end{equation}
where $\lambda_i^0$ is the laboratory (vacuum) wavelength of the transition.
This formula can be written in term of redshift
$z_i=\lambda_i/\lambda_i^0-1$ as
\begin{equation}
z_i=z_{abs}+b \!\cdot \! \! K_i \, ,
\label{korr}
\end{equation}
where $b=(1+z_{abs})\cdot\Delta\mu/\mu$. In reality, $z_i$ is measured
with some uncertainty which is caused by statistical errors of
the astronomical measurements $\lambda_i$ and also errors of the
laboratory measurements of the wavelengths $\lambda_i^0$.
Thus, a linear regression analysis yields $z_{abs}$ and $b$ (consequently
$\Delta\mu/\mu$) with their statistical significance.

\section*{Observations and Results of Analysis}
\hspace{1cm}
To measure the possible variation of
$\mu$ we use high resolution spectra (FWHM $\approx$ 7~km/s) of quasars
obtained with UVES at the 8.2-m ESO VLT Kueyen telescope.
Two H$_2$ absorption systems were analysed at $z_{\rm abs}=2.3377$ in the
spectrum of Q~1232+082 (Petitjean et al. 2000), and
at $z_{\rm abs}=3.0249$ in the spectrum of Q~0347-382
(UVES commissioning data, see D'Odorico et al. 2001).\par

\vspace{0.7cm} \hspace{-1.2cm}{\it Absorption system of H$_2$ at
z=2.3377 in the spectrum of Q~1232+082}\\
More than 50 lines of molecular hydrogen (with S/N $\sim$ 10-14)
are identified in the range 3400-3800 \AA. We have carefully
selected lines that are isolated, unsaturated and unblended. In
this system only 12 lines meet all these conditions (see Table~1).
The accuracy on the observed wavelengths $\lambda_i$ calculated in
accordance with Eq.~(A14) from Bohlin et al. (1983) which takes
into account the number of points within the profile of the
spectral line, the spectral resolution and the S/N ratio. The
average uncertainty of the determination of the line centers is
$\sigma(\lambda_i) \sim 5$~m\AA. For the laboratory wavelengths we
have used two independent sets of data: $\lambda_i^0 \; [M]$ from
Morton \& Dinerstein, 1976; and $\lambda_i^0 \; [A]$ from Abgrall
et al., 1993 (see, also Roncin \& Launay, 1994). Results of the
linear regression analysis of $z_i$-to-$K_i$ are shown on Fig.~1
for both sets of laboratory wavelengths. They are the following:
$\Delta \mu / \mu = (14.4\pm11.4)\times 10^{-5}$~[A], and $\Delta
\mu / \mu =
 (13.2\pm7.4)\times 10^{-5}$~[M].

\par \vspace{0.7cm}
\hspace{-1.2cm}{\it Absorption system of H$_2$ at z=3.0249 in the
spectrum of Q~0347-382}\\
For the first time, this H$_2$ system was found and investigated
by Levshakov et al., 2001. More than 80 lines of molecular
hydrogen (with S/N $\sim$ from 10 to 40) can be identified in the
wavelength range 3600-4600 \AA. We have reanalysed this spectrum
independently. For our analysis 18 lines of H$_2$ were selected
which satisfied to the conditions mentioned above. The average
uncertainty of the determination of the line centers is
$\sigma(\lambda_i) \sim 10$~m\AA. Parameters of the lines are
presented in Table~2. The results of the linear regression
analysis of $z_i$-to-$K_i$ are shown on Fig.~2 for both sets of
laboratory wavelengths. They are the following: $\Delta \mu / \mu
= (5.8\pm3.4)\times 10^{-5}$~[A], and $\Delta \mu / \mu =
(12.2\pm7.3)\times 10^{-5}$~[M]. It should be mention that three
points on the bottom panel of Fig.~2 depart from the regression
line more than $3\sigma$. Two of them corresponding to L~9-0~R(1)
and W~1-0~R(1) transitions have the laboratory wavelengths marked
by Morton \& Dinerstein (1976) as a blended line and a line under
weak continuum. The third point corresponding to W~3-0~Q(1)
transition has deviations on the both panels. It may be a result
of
 undetectable blending in the quasar spectrum. We do not reject
these points because all of them satisfy to the above formulated
conditions for the line selection
from quasar spectra.

\vspace{0.7cm}
\hspace{-1.2cm}{\it Combined Analysis}\\
The combined analysis of the H$_2$ lines from the two systems discussed above
allows us to increase the statistical significance of the estimate
because of increasing the number of lines involved and,
what is more essential, broadening the interval of K-values.
The results of linear regression analysis of $\zeta_i$ as a function of
$K_i$ for all 30 lines from both systems are shown in Fig.~3. Here $\zeta_i$
is the reduced value of the line redshift:
\begin{equation}\zeta_i = \frac{z_i-\overline{z}}{1+\overline{z}} \, ,
\end{equation}
where $\, \overline{z} \,$ is $z_{abs}$ corresponding to the absorption system
and a particular set of laboratory wavelengths.

As a result of the combined analysis we obtained the following
estimates (for two different sets of laboratory wavelengths):
\[\Delta \mu / \mu = (5.7\pm 3.8)\times10^{-5} \;\;\; [A] \, ,
\]\[\Delta \mu / \mu = (12.5\pm 4.5)\times10^{-5} \;\;\; [M] .\]
The statistical uncertainties of the laboratory wavelengths are about
$1.5\,$m\AA~ corresponding to an error of
$\Delta \mu / \mu$ about $2\times 10^{-5}$ that is
in agreement with the errors found from the regression.

\section*{Conclusions}
\hspace{1cm}
The above results may be considered as a glimpse on possible cosmological
variation of $\mu$. Additional measurements are necessary to ascertain
the conclusion.

In any case, we have obtained the most stringent estimate on
a possible cosmological variation of $\mu$ between redshift zero and
redshifts 2--3.

In order to improve the result, it is necessary to measure more H$_2$
absorption systems at high redshift.
The most suitable quasars for such analysis are PKS~0528-250, Q~0347-382,
and Q~1232+082. Their observations with high resolution (FWHM $\sim7$ km/s)
and high S/N ratio ($>30$) will strengthen the conclusions.

In addition,
better accuracy of laboratory wavelengths is also desirable because the
contribution of laboratory statistical errors are comparable to the
statistical errors of astronomical measurements.
\par\bigskip\noindent
Acknowledgments:{\it  ~The observations have been obtained with
UVES mounted on the \mbox{8.2-m} KUEYEN telescope operated by the
European Southern Observatory at Parana, Chili. The authors thank
C.~Ledoux for primary reduction of the spectra and A.~Potekhin for
useful discussion. A.~Ivanchik and
 D.~Varshalovich are grateful for
the support by the RFBR
 (01-02-06098, 99-02-18232).
A.~Ivanchik is grateful for the
 opportunity to visit the IAP~CNRS. }

\newpage
\section*{References}
{\fontsize{11pt}{27mm}\selectfont
Abgrall~H., Roueff~E., Launay~F., Roncin~J.-Y., Subtil~J.-L. // Astron. Astrophys. Suppl. Ser., 1993, V. 101, P. 273. \\
Bohlin~R.C., Hill~J.K., Jenkins~E.B., Savage~B.D., Snow~Jr.~T.P., Spitzer~Jr.~L., York~D.G. // Astrophys. J. Suppl. Ser., 1983, V. 51, P. 277. \\
Calmet~X., Fritzsch~H. // 2001, /hep-ph/0112110. \\
Damour~T., Polyakov~A.M. // Nucl. Phys., 1994, B423, P. 532. \\
D'Odorico~S., Dessauges-Zavadsky~M., Molaro~P. // Astron. Astrophys., 2001, V. 368, P. L1. \\
Gasser~J., Leutwyler~H. // Physics Reports, 1982, V. 87, No. 3, P. 77. \\
Ivanchik~A.V., Potekhin~A.Y., Varshalovich~D.A. // Astron. Astrophys., 1999, V. 343, P. 439. \\
Levshakov~S.A., Dessauges-Zavadsky~M., D'Odorico~S., Molaro~P. // Astrophys. J., 2002, V. 565, in press. \\
Mohr~P.J, Taylor~B.N. // Reviews of Modern Physics, 2000, V. 72, No. 2, P. 351. \\
Morton~D.C., Dinerstein~H.L. // Astrophys. J., 1976, V. 204, P. 1. \\
Petitjean~P., Srianand~R., Ledoux~C. // 2000, /astro-ph/0011437. \\
Potekhin~A.Y., Ivanchik~A.V., Varshalovich~D.A., Lanzetta~K.M., Baldwin~J.A., Williger~G.M., Carswell~R.F. // Astrophys. J., 1998, V. 505, P. 523. \\
Ronchin~J.-Y., Launay~F., // Journal Phys. and Chem. Reference Data, 1994, No. 4. \\
Varshalovich~D.A., Levshakov~S.A. // JETP Letters, 1993, V. 58, P. 231. \\
Varshalovich~D.A., Potekhin~A.Y. // Space Science Rev., 1995, V. 74, P. 259. \\
Vysotsky~M.I., Novikov~V.A., Okun'~L.B., Rozanov~A.N. // Physics-Uspekhi, 1996, V. 166, No. 5, P. 539. \\
Webb~J.K., Murphy~M.T., Flambaum~V.V., Dzuba~V.A., Barrow~J.D., Churchill~C.W., Prochaska~J.X., Wolfe~A.M. // Phys. Rev. Lett., 2001, V. 87, P. 091301. \\
}

\newpage\pagestyle{empty}
\begin{table}[t]
\caption{H$_2$ lines of absorption system at $z_{abs}=2.3377$ in Q~1232+082
spectrum}     \[
\begin{array}{rrrrcr}
\hline
\hline
\noalign{\smallskip}
\mbox{Lines~~~~} & \;\;\;\; \lambda_i^0, \, \mbox{\AA} \; [M] & \;\;\;\;
 \lambda_i^0, \, \mbox{\AA} \; [A] & \;\;\;\; \lambda_i, \, \mbox{\AA} \; &
\sigma(\lambda_i), \, \mbox{\AA} & K_{i} \;\;\;\; \\
\noalign{\smallskip}
\hline
\noalign{\smallskip}
\mbox{L~0-0~P(3)} & 1115.896 &  1115.895 &  3724.543  & 0.005 &  -0.01479 \\
\mbox{L~0-0~R(3)} & 1112.584 &  1112.583 &  3713.480  & 0.005 &  -0.01178 \\
\mbox{L~0-0~P(2)} & 1112.495 &  1112.459 &  3713.179  & 0.005 &  -0.01170 \\
\mbox{L~1-0~P(4)} & 1104.084 &  1104.084 &  3685.093  & 0.009 &  -0.01154 \\
\mbox{L~2-0~P(3)} & 1084.559 &  1084.562 &  3619.947  & 0.005 &  -0.00098 \\
\mbox{L~3-0~P(4)} & 1074.313 &  1074.314 &  3585.740  & 0.007 &   0.00122 \\
\mbox{L~3-0~R(4)} & 1070.898 &  1070.899 &  3574.344  & 0.006 &   0.00439 \\
\mbox{L~3-0~P(3)} & 1070.142 &  1070.138 &  3571.834  & 0.005 &   0.00511 \\
\mbox{L~3-0~R(3)} & 1067.478 &  1067.474 &  3562.948  & 0.005 &   0.00758 \\
\mbox{L~3-0~P(2)} & 1066.901 &  1066.899 &  3561.002  & 0.005 &   0.00812 \\
\mbox{L~4-0~P(4)} & 1060.580 &  1060.580 &  3539.908  & 0.005 &   0.00685 \\
\mbox{L~4-0~P(2)} & 1053.281 &  1053.283 &  3515.556  & 0.005 &   0.01369 \\
\noalign{\smallskip}
\hline
\end{array}      \]
\end{table}
\begin{table}[t]
\caption{H$_2$ lines of absorption system at $z_{abs}=3.0249$ in Q~0347-382
spectrum}     \[
\begin{array}{rrrrcr}
\hline
\hline
\noalign{\smallskip}
\mbox{Lines~~~~} & \;\;\;\; \lambda_i^0, \, \mbox{\AA} \; [M] & \;\;\;\;
\lambda_i^0, \, \mbox{\AA} \; [A] & \;\;\;\; \lambda_i, \, \mbox{\AA} \; &
\sigma(\lambda_i), \, \mbox{\AA} & K_{i} \;\;\;\; \\
\noalign{\smallskip}
\hline
\noalign{\smallskip}
\mbox{L~~2-0~R(1)} & 1077.698 & 1077.697 & 4337.614 & 0.010 &  0.00535  \\
\mbox{L~~3-0~R(2)} & 1064.995 & 1064.993 & 4286.483 & 0.015 &  0.00989  \\
\mbox{L~~3-0~P(1)} & 1064.606 & 1064.606 & 4284.924 & 0.006 &  0.01026  \\
\mbox{L~~3-0~R(1)} & 1063.460 & 1063.460 & 4280.313 & 0.010 &  0.01132  \\
\mbox{L~~4-0~R(3)} & 1053.976 & 1053.977 & 4242.144 & 0.010 &  0.01304  \\
\mbox{L~~4-0~R(2)} & 1051.497 & 1051.498 & 4232.175 & 0.020 &  0.01536  \\
\mbox{L~~6-0~R(3)} & 1028.986 & 1028.983 & 4141.571 & 0.015 &  0.02262  \\
\mbox{L~~7-0~R(1)} & 1013.434 & 1013.436 & 4078.977 & 0.007 &  0.03062  \\
\mbox{W~~0-0~Q(2)} & 1010.941 & 1010.938 & 4068.911 & 0.010 & -0.00686  \\
\mbox{W~~0-0~R(2)} & 1009.030 & 1009.023 & 4061.215 & 0.015 & -0.00503  \\
\mbox{L~~9-0~R(1)} &  992.022 &  992.013 & 3992.754 & 0.010 &  0.03796  \\
\mbox{W~~1-0~Q(2)} &  987.978 &  987.974 & 3976.492 & 0.010 &  0.00394  \\
\mbox{W~~1-0~R(1)} &  985.651 &  985.636 & 3967.087 & 0.007 &  0.00626  \\
\mbox{L~10-0~P(1)} &  982.834 &  982.834 & 3955.814 & 0.010 &  0.04053  \\
\mbox{L~12-0~R(3)} &  967.674 &  967.675 & 3894.798 & 0.008 &  0.04386  \\
\mbox{W~~2-0~Q(2)} &  967.278 &  967.279 & 3893.194 & 0.010 &  0.01301  \\
\mbox{W~~2-0~Q(1)} &  966.097 &  966.094 & 3888.423 & 0.007 &  0.01423  \\
\mbox{W~~3-0~Q(1)} &  947.425 &  947.422 & 3813.255 & 0.008 &  0.02176  \\
\noalign{\smallskip}
\hline      \end{array}      \]
\end{table}

\newpage

\begin{figure}[h]
 \centering
   \includegraphics[height=190mm,bb=70 70 550 780,clip]{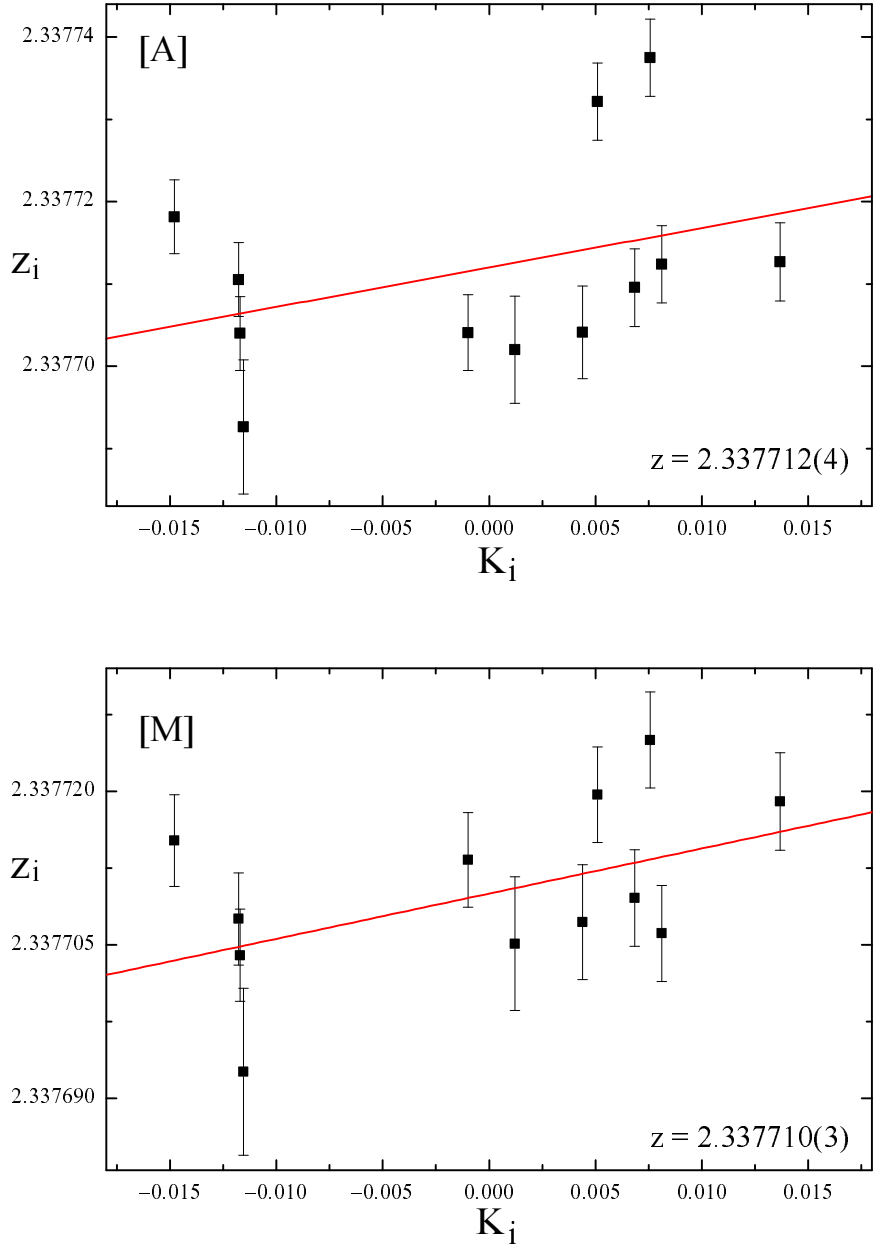}
   \fontsize{10pt}{10pt}\selectfont
   \caption{\large ~~Results of $z_i$-to-$K_i$ regression analysis for H$_2$
   lines
 at $z_{abs}=2.3377$ in the Q~1232+082 spectrum.
   Upper panel: laboratory wavelengths are taken from Abgrall et al. (1993).
   Bottom panel: laboratory wavelengths are taken
   from Morton \& Dinerstein (1976).}
   \end{figure}

\newpage

\begin{figure}[h]
 \centering
   \includegraphics[height=190mm,bb=70 70 550 780,clip]{figure2.ps}
   \fontsize{10pt}{10pt}\selectfont
   \caption{\large ~~Results of $z_i$-to-$K_i$ regression analysis for H$_2$
   lines
 at $z_{abs}=3.0249$ in the \mbox{Q~0347-382} spectrum.
   Upper panel: laboratory wavelengths are taken from Abgrall et al. (1993).
   Bottom panel: laboratory wavelengths are taken
   from Morton \& Dinerstein (1976).
}
   \end{figure}

\newpage

\begin{figure}[h]
 \centering
   \includegraphics[height=190mm,bb=70 70 550 780,clip]{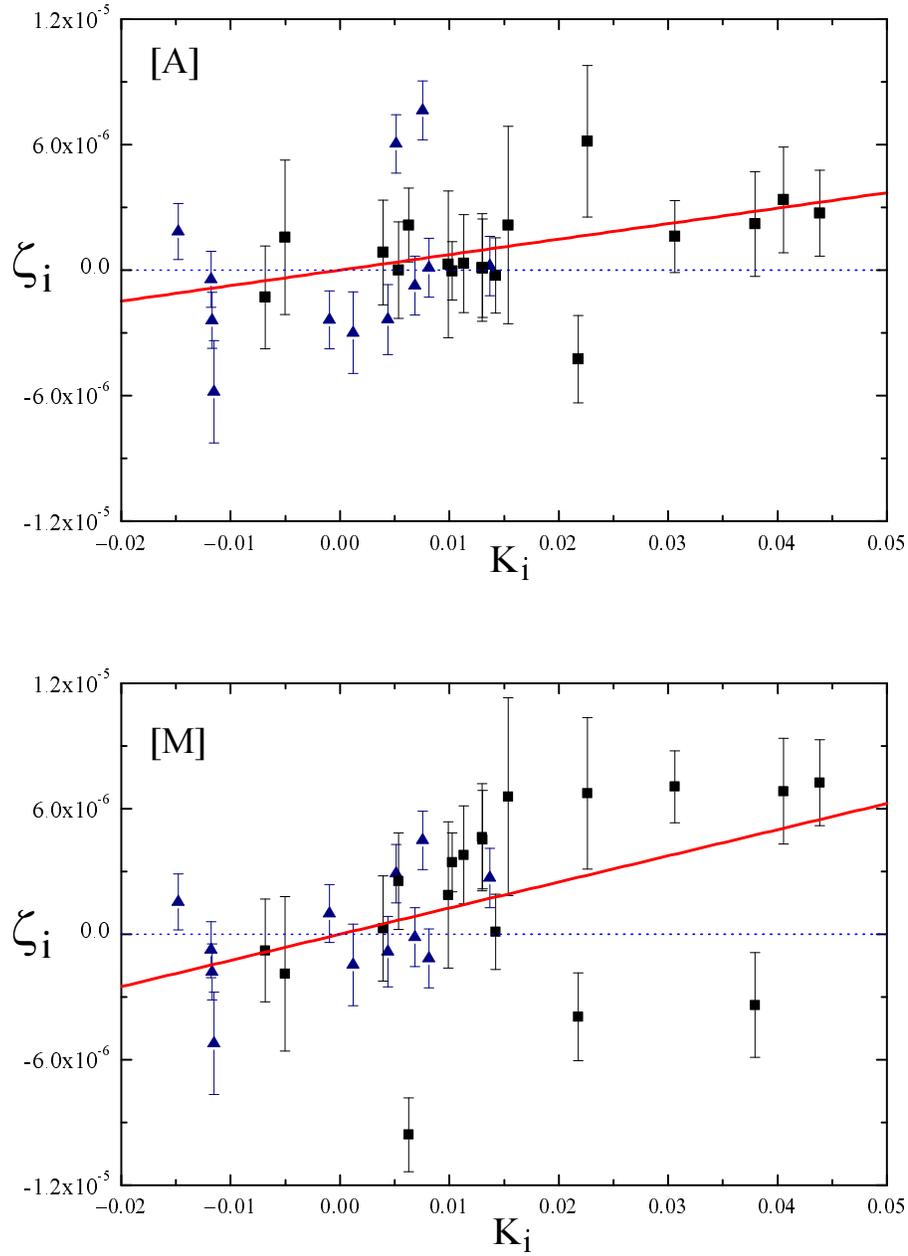}
   \fontsize{10pt}{10pt}\selectfont
   \caption{\large ~~Combined regression analysis of $\zeta_i$-to-$K_i$
   for the systems
 at $z_{abs}=2.3377$ (triangles) and $3.0249$ (squares).
   Upper panel: laboratory wavelengths are taken from Abgrall et al. (1993).
   Bottom panel: laboratory wavelengths are taken
   from Morton \& Dinerstein (1976).
}
   \end{figure}

\end{document}